\documentclass[conference]{IEEEtran}
\IEEEoverridecommandlockouts
\usepackage{cite}
\usepackage{amsmath,amssymb,amsfonts}
\usepackage{algorithmic}
\usepackage{graphicx}
\graphicspath{{figures/}}
\usepackage{textcomp}
\usepackage{xcolor}
\usepackage{booktabs}
\usepackage{enumitem}
\usepackage{comment}
\usepackage{balance}
\usepackage{subcaption}
\usepackage[hidelinks]{hyperref}

\begin{document}
\newcommand\todo[1]{\textcolor{red}{#1}}
\newcommand\new[1]{#1}

\title{Evaluating Emerging CXL-enabled Memory Pooling for HPC Systems
}

\author{\IEEEauthorblockN{Jacob Wahlgren}
\IEEEauthorblockA{\textit{Department of Computer Science} \\
\textit{KTH Royal Institute of Technology}\\
Stockholm, Sweden \\
jacobwah@kth.se}
\and
\IEEEauthorblockN{Maya Gokhale}
\IEEEauthorblockA{\textit{CASC} \\
\textit{Lawrence Livermore National Laboratory}\\
Livermore, USA \\
gokhale2@llnl.gov}
\and
\IEEEauthorblockN{Ivy B. Peng}
\IEEEauthorblockA{\textit{Department of Computer Science} \\
\textit{KTH Royal Institute of Technology}\\
Stockholm, Sweden \\
bopeng@kth.se}
}

\maketitle

\begin{abstract}
Current HPC systems provide memory resources that are statically configured and tightly coupled with compute nodes. However, workloads on HPC systems are evolving. Diverse workloads lead to a need for configurable memory resources to achieve high performance and utilization. In this study, we evaluate a memory subsystem design leveraging CXL-enabled memory pooling. Two promising use cases of composable memory subsystems are studied -- fine-grained capacity provisioning and scalable bandwidth provisioning. We developed an emulator to explore the performance impact of various memory compositions. We also provide a profiler to identify the memory usage patterns in applications and their optimization opportunities. Seven scientific and six graph applications are evaluated on various emulated memory configurations. Three out of seven scientific applications had less than 10\% performance impact when the pooled memory backed 75\% of their memory footprint. The results also show that a dynamically configured high-bandwidth system can effectively support bandwidth-intensive unstructured mesh-based applications like OpenFOAM. Finally, we identify interference through shared memory pools as a practical challenge for adoption on HPC systems.
\end{abstract}


\section{Introduction}
As high-performance computing (HPC) systems enter the exascale era, hardware specialization and resource utilization pose a continued challenge. Today, HPC systems are equipped with tightly coupled memory, compute, and storage resources within the node boundary and rely on resource over-provisioning to support characteristically diverse workloads. Such static coarse-grained provisioning simplifies resource management.
However, recent studies on large-scale supercomputers indicate that node-level memory utilization can be as low as 15\%~\cite{peng2020memory,michelogiannakis2022case}. One reason is that memory bandwidth and capacity are tightly coupled in current HPC systems. Consequently, bandwidth-intensive jobs like computational fluid dynamics (CFD) codes often need to request large memory capacity to meet their bandwidth requirements, resulting in significant memory under-utilization.
Meanwhile, the diversity of workloads on HPC systems also increases when more machine learning and data analytics components are integrated into workflows. They face different limiting factors in memory subsystems. Finally, the increased heterogeneity in HPC systems further complicates resource utilization. Specialized accelerators like GPUs are becoming ubiquitous, and these heterogeneous compute units require adequately provisioned memory resources to sustain their performance.

Disaggregating memory from compute and providing on-demand memory resources from memory pooling is a common approach for improving memory utilization. Essentially, memory pooling mitigates memory imbalance and provides temporary memory expansion to address memory underutilization. Many works have explored pooling stranded memory resources to support other jobs~\cite{Lim2012, peng2020memory}. They use network-attached memory as memory pools and provide software extensions to paging management to ease programming efforts. However, these solutions often bring noticeable performance degradation due to overheads of swapping and network latency. Previous works in memory pooling and disaggregation focus on cloud and data center workloads and their characteristics~\new{\cite{Li2022,Maruf2022,Gouk2022,Lim2012,lagar2019software,agarwal2017thermostat,park2022scaling,dulloor2016data}}.

Recently, scalable hardware-supported memory disaggregation and pooling have become possible with the Compute Express Link (CXL) standard~\cite{CXL3Spec}. CXL is an open standard for interconnecting processors, accelerators, and memory. Hardware conforming to the CXL standard provides low-latency, high-bandwidth data access transparently to application codes. Several works have explored CXL-based memory for data center and cloud workloads~\cite{Li2022, Gouk2022, Maruf2022, park2022scaling}. However, a comprehensive understanding of CXL-enabled memory pooling for HPC systems and workloads is still missing.

\begin{figure}
\centering
\includegraphics[width=\linewidth]{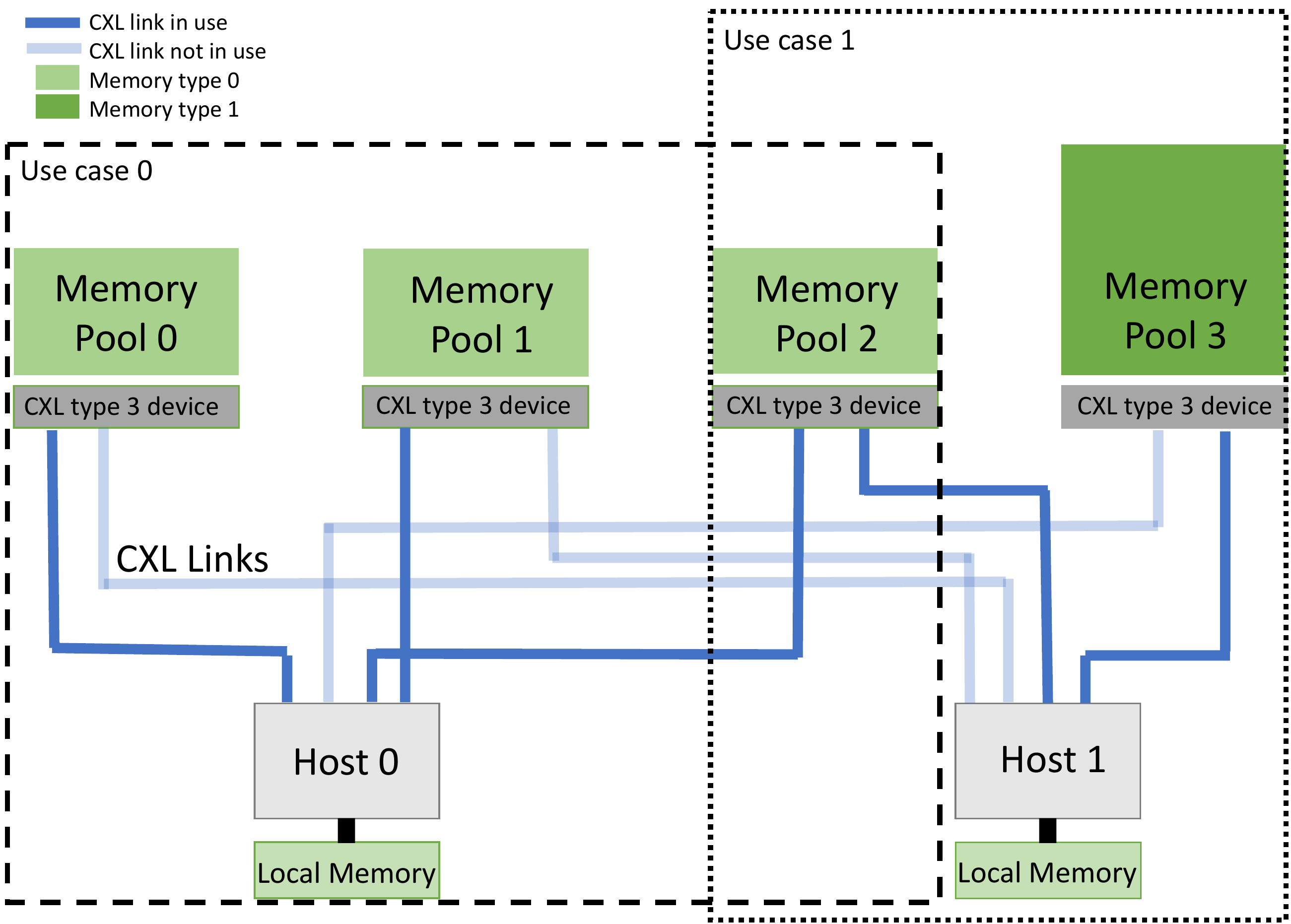}
\caption{An potential composable memory system design. CXL enables multiple memory organizations on one system. Each CXL link connects a host to a memory pool. One host can be connected with CXL links to multiple memory pools. Different memory pools may be backed by different memory types. Use Case 0 illustrates that Host 0 is dynamically configured with a memory subsystem of scaled bandwidth (3 links). Use Case 1 illustrates that Host 1 is configured with a heterogeneous memory system composed of two memory types.}
\label{fig:arch}
\end{figure}

In this work, we focus on the \textit{CXL.mem} protocol and CXL type 3 devices to investigate their feasibility of implementing composable memory subsystems on future HPC systems. We show that a single system design illustrated in Figure~\ref{fig:arch} can be dynamically configured into multiple memory subsystems to match the requirements of characteristically different workloads, e.g., the seven dwarfs in scientific computing~\cite{Asanovic2006}. In particular, the dynamic configurations separate capacity provision from bandwidth provision, which is impossible on current systems. As the CXL standard supports native load/store instructions to access CXL-connected memory, the changes to the memory subsystem are transparent to applications.

Due to the lack of real hardware, we provide an emulator for fast exploration of applications on various memory configurations. Our tool also profiles dynamic memory usage in capacity, bandwidth, and page hotness, to understand optimization options. We evaluated two use cases of the configurable CXL-based memory system. First, we evaluate configurations that use the pooled memory to back an increased percentage of memory footprint in HPC and graph workloads. Then, we characterized their performance impacts and identified three classes of workloads based on their sensitivity to memory pooling. Two out of the three classes can leverage CXL-based memory pooling with a low impact on performance. For the third class, we also evaluate a second system configuration that scales bandwidth using an increased number of CXL links between one host and memory pools. The results show that this configuration may provide a cost-effective high-bandwidth system for bandwidth-bound workloads such as unstructured mesh-based simulations.
 
Several practical challenges in adopting CXL-based memory pooling on HPC systems have been identified. Extensive works have looked into data placement and memory management on multi-tier and heterogeneous memory systems~\cite{dulloor2016data,agarwal2017thermostat,Maruf2022}. Instead, we evaluated the effect of sharing a memory pool among similar and unrelated programs running on multiple hosts. Our results show that system-level support is necessary to detect and mitigate interference from applications with conflicting memory usage patterns on a shared pool. In summary, we made the following contributions in this work:

\begin{itemize}[leftmargin=*]
    \item We demonstrate a CXL-based memory system design supporting composable memory capacity and bandwidth.
    \item We develop tools for emulating applications on various configurations of the memory system.
    \item Five out of seven scientific applications can sustain performance with 75\% memory footprint backed by the memory pool.
    \item A high-bandwidth configuration could be a cost-effective solution to support bandwidth-intensive workloads like OpenFOAM.
    \item System-level coordination is needed to mitigate interference through memory pooling among workloads with conflicting usage patterns.
\end{itemize}

\section{Background}
Most HPC systems employ a server-centric design, where each compute node comprises a fixed amount of compute and memory resources. Resources are underutilized when a workload uses compute and memory differently from the fixed resource configuration. One promising solution to address underutilization is memory and compute disaggregation. On disaggregated memory systems, compute resources are loosely coupled with a memory pool that provides on-demand memory resources to workloads. As compute and memory resources can be independently provisioned, the overall utilization improves. Also, the memory pool can be maintained and upgraded independently from compute resources, potentially reducing the total cost of ownership (TCO).

Several cache-coherent interconnect protocols can be used to enable memory and compute disaggregation. For instance, IBM implemented the OpenCAPI specification in Power9 and Power10 processors to enable coherent memory sharing among hosts and accelerators. Open Memory Interface (OMI) is an OpenCAPI subset providing a technology-agnostic memory interface. Both Gen-Z and CXL are memory-semantic protocols that support direct load and store to fabric-attached memory. NVLink from Nvidia and Infinity Fabric from AMD are both proprietary accelerator interconnects that aim to provide tighter memory sharing between GPU and CPU. As both OpenCAPI and Gen-Z joined the CXL consortium in 2022, CXL has become the most widely endorsed open interconnect standard. 

\subsection{CXL Standard}
The CXL standard defines three types of devices -- type~1 (accelerator with cache), type~2 (accelerator with memory), and type~3 (memory expander). Overall, CXL enables a unified coherent memory space among the host and CXL-attached devices, e.g., accelerators. Host and devices can communicate in three protocols known as \textit{CXL.io}, \textit{CXL.cache}, and \textit{CXL.mem}. Memory access is supported through native load and store instructions, which translate to \textit{CXL.mem} and optionally \textit{CXL.cache} transactions.

The latest version CXL~3.0~\cite{CXL3Spec} is based on PCIe~6.0 and supports up to 64~GT/s data rate, e.g., 256 GB/s raw bidirectional bandwidth (x16). The previous CXL~2.0 is based on PCIe~5.0 and supports a peak rate of 32~GT/s. Recommended latency targets for a simple type 3 device are 80~ns for reads and 40~ns for writes~\cite{CXL3Spec}. The recommended link layer latency at high load is 65~ns over PCIe~5.0 and is expected to be lower over PCIe~6.0. Additional latency may be introduced if there are intermediate switches, e.g., CXL switches. In this work, we target type 3 devices operating on the \textit{CXL.mem} protocol.

\subsection{A CXL-enabled Composable Memory Subsystem}
Figure~\ref{fig:arch} illustrates a composable memory subsystem design that can be achieved through CXL, similar to~\cite{Li2022}. In this example, each CXL type 3 device acts as a memory pool providing memory resources to two hosts, which may run different MPI ranks of a job or even unrelated jobs.
A process on Host 0 could have a virtual memory space backed by four physical memory devices -- the local memory directly attached to the host and three type 3 CXL memory pools (i.e., Pool 0-2). Each device is connected to a memory pool (interchangeably referred to as a memory server) by a separate CXL link. A memory pool serves multiple hosts, e.g., Memory Pool 2 provides memory resources to both Host 0 and Host 1.

Composability is one main motivation for using a CXL-enabled memory subsystem. As illustrated in Figure~\ref{fig:arch}, a job could be scheduled to a set of hosts, where the connection between a host and memory pools can be dynamically enabled to construct a memory subsystem that meets the job's memory usage. For instance, applications that can leverage different memory types can compose a heterogeneous memory system on demand. The main challenge in supporting memory composability comes from mapping each application's memory usage to the available physical resources.

Memory pooling can be supported at different granularity levels. The lowest overhead for management is static allocation per job. Each job could request a fixed portion of memory resources from memory pools at start time. Since a memory pool can be shared by multiple jobs or ranks, this approach may increase utilization as compared to the current server-centric systems.
To increase utilization further, memory resources could scale up and down during a job to match the dynamic memory usage. However, the overhead of changing the composition of a memory system, i.e., adding physical memory at runtime, may be expensive and require extensive monitoring support. 

Performance optimizations on disaggregated memory are similar to existing heterogeneous and multi-tier memory. On such systems, the virtual memory space is backed by multiple memory types, such as DRAM, HBM, and Persistent Memory. Many works have previously focused on extending NUMA systems to improve data placement and migration between different memory tiers. However, CXL-enabled systems pose new challenges. For instance, our characterization study highlighted how unrelated jobs and classical MPI-like applications could experience different contention and interference on shared memory pools.

\section{Methodology}
In this section, we propose an emulator for evaluating different memory configurations on the system architecture introduced in Fig.~\ref{fig:arch}. To analyze memory usage behaviors and identify optimization strategies, we also propose a profiler that identifies dynamic memory usage patterns of applications.

\subsection{Profiler}
To provide a fast and accurate estimation of whether a given application could leverage remote CXL-enabled memory, we need to assess several memory utilization metrics, including capacity, bandwidth, and dynamic usage. Thus, we provide a profiling tool that quantifies memory usage metrics such as working set size and accessed pages in a parameterized profiling interval. The tool provides a temporal profile of these memory usage dynamics, which could reveal more insights than a summarized metric per job as in existing works.

\subsection{Emulator}
To enable fast exploration of potential applications and the impact of CXL memory servers on their performance, we design and implement an emulator that runs on existing NUMA-enabled hardware. Our tool leverages NUMA machines to emulate a CXL-enabled memory subsystem with configurable local memory and remote memory capacity. 

\subsection{Implementation}

The profiler and emulator are implemented in ${\sim}600$ lines of C code and will be released as open-source. It is compatible with Linux kernel 4.14 and above. The emulator relies on the \verb|libnuma| library.

To capture memory usage metrics, the profiler and emulator use files in the \verb|proc| file system, in the \verb|/proc/pid| sub-tree. This provides a portable method requiring no special privileges.
Metrics like the resident set size and the number of referenced memory pages are read from \verb|smaps_rollup|, and the number of pages in each NUMA domain is read from \verb|numa_maps|. Reading from these files causes the kernel to scan the memory areas of a process and return the requested metrics.
The profiler can also reset the referenced status of all pages by writing to \verb|clear_refs|.

To capture the temporal memory usage of a process, the profiler operates in a timer mode, where it collects measurements and resets the page status with a configurable frequency. To capture summarized metrics such as the number of cold pages, the profiler operates in interrupt mode utilizing Linux job control features. We modify the benchmarks to raise a \verb|SIGSTOP| signal which pauses the process at the start and end of the timed region. Pausing the process wakes the profiler with a \verb|SIGCHLD| signal. The profiler collects the measurements and resets the page status, and then resumes the application by sending a \verb|SIGCONT| signal. For cases when it is not possible to change the code of a benchmark, we also provide an output-interrupt mode, which monitors the standard output of the processes and pauses it when a configurable pattern matches the output. An \verb|epoll_wait| loop is used to wait for events like timers, signals, and output.

In the emulator, a combination of NUMA policies and locked memory is used to configure an emulated CXL memory system. The application is limited to run on NUMA node~0 by using \verb|numa_run_on_node()|. We let the memory in node~0 represent the local/near memory, and the memory in other NUMA nodes represent remote/far memory that is accessed through CXL. The default allocation policy will use the local node until it is full, and then start using the other nodes. To emulate a system with a certain ratio of local to remote memory, we lock a suitable amount of local memory using \verb|mlock()| to force the allocator to use remote memory. To emulate a system with only remote memory, we instead set the allocation policy to only use node~1 with \verb|numa_set_membind()|.

The sampling overhead is tunable, based on the user-defined sampling frequency. With a frequency of 1 s, the profiler has a sampling overhead of 3\%, and the emulator has a sampling overhead of 2\% compared to an equivalent \verb|numactl| execution. 
The emulator has an additional initialization time proportional to the amount of local memory when using local--remote ratios other than 0 or 1 due to locking a large amount of memory.

Workloads using MPI consist of multiple processes. In the profiler, we choose to sample rank 0. In the emulator, we sample all the processes independently and aggregate the results as a post-processing step.

\subsection{\new{Evaluation Workflow}}

We propose the following workflow to evaluate the performance impact of CXL-enabled memory pooling on an application. In the later sections, this workflow is applied to a diverse set of HPC and graph applications.
\begin{enumerate}
\item A representative input problem is chosen.
\item\label{it:memuse} The dynamic memory usage of the application is profiled. If the memory usage varies significantly during the execution, memory pools may need to be dynamically configured during runtime. On the other hand, if the memory usage is stable, memory pools can be configured at job startup.
\item The dynamic memory access pattern of the application is profiled. If some part of the application's working set is seldom or never accessed, that part is a good candidate to store in the slower memory pool without impacting the performance.
\item Using the emulator, the application is executed with different amounts of pooled memory. With no pooled memory as a baseline, the impact on the execution time of increasing the amount of pooled memory is analyzed. The application is categorized as largely insensitive to increased pooled memory, moderately sensitive, or highly sensitive. Insensitive applications can take advantage of pooled memory without modification. Moderately sensitive applications are candidates for optimization efforts.
\item If the application was shown to be highly sensitive in the previous step, it may instead benefit from increased bandwidth enabled by spreading the working set across multiple memory pools. The execution time of the application is evaluated when increasing the number of emulated CXL links.
\item To evaluate the impact of interference when sharing a memory pool, several concurrent instances of the application are executed using the same emulated memory pool. The application is also executed together with other unrelated workloads using the same emulated memory pool. This step can indicate scheduling constraints required to maintain performance in a multi-node or multi-user system.
\end{enumerate}

\section{Experimental Setup}\label{sec:setup}
\begin{table}[bt]
	\caption{Evaluated workloads.\label{tab:app}}
	\resizebox{\linewidth}{!}{
	\centering
	\begin{tabular}{ll}
	\toprule
    \textbf{HPC Applications} 			&\textbf{Input Problem}\\\midrule
    BLAS                    &20,000x20,000 randomized matrices\\
    SuperLU                 &sparse matrix with 1.3M non-zeroes (SiO~\cite{Davis2011})\\
    NPB-FT                  &3D PDE on $512^3$ grid (class C)\\
    SPLASH-BARNES           &17M particles\\
	Hypre                   &2D CDR problem, 56M grid points (ex4)\\
    OpenFOAM                &HPC motorbike benchmark, 8.5M cells\\
    XSBench                 &unionized history on "large" problem, 2M particles\\
    \midrule
	\textbf{Graph Applications} 	&\textbf{Description}\\\midrule
	BFS                &breadth-first search\\
	BC                 &betweeness centrality\\
	Radii              &eccentricity estimation\\
	Components		   &connected components\\
	PageRank           &website ranking\\
	Bellman-Ford       &weighted shortest paths\\ \bottomrule
 \end{tabular}}
\end{table}

In this section, we describe our emulation platforms and the evaluated workloads.

\subsection{Testbeds}
We use two NUMA testbeds for emulating potential configurations of the CXL-enabled memory system as introduced in Fig.~\ref{fig:arch}. The first test bed has dual-socket Intel Xeon E5-2690V4 processors connected by two QPI links, configured as two NUMA domains. It has 64 GB memory capacity per NUMA node and a total of 128 GB memory. The second testbed has one AMD EPYC 7742 processor with four NUMA domains. It consists of four core complexes, each with a separate memory controller providing 33~GB/s bandwidth respectively. This testbed has 32 GB memory capacity per NUMA node and a total of 128 GB memory.

\subsection{Workloads}
In this study, our focus is on HPC workloads. For contrast, we also include representative graph workloads that might benefit from the capacity scaling of memory pools. For graph workloads, we use six applications from the Ligra~\cite{shun2013ligra} graph processing framework on an RMAT scale 24 input graph.
Additionally, to measure memory bandwidth we use STREAM triad, and to measure memory latency we use a pointer-chasing benchmark~\cite{Foyer2021}.
For comprehensive coverage of HPC workloads, we select seven applications from the ``Seven Dwarfs"~\cite{Asanovic2006} of major numerical algorithms. The dwarfs and their corresponding HPC applications are listed below.

\begin{itemize}[leftmargin=*]
\item Dense Linear Algebra includes vector and matrix operations that are often highly optimized for cache locality and common in compute-intensive kernels. In this work, we use the generalized matrix-matrix product (\verb|DGEMM|) from BLAS~\cite{Lawson1979}, a level 3 operation with high operational intensity. On the Intel platform, we use the MKL implementation and on the AMD platform, we use the Cray-libsci implementation.

\item Sparse Linear Algebra computes data stored in compressed formats and features indirect memory access. We use sparse LU factorization in SuperLU~\cite{Li2005}.
\item Spectral Methods rely on the fast Fourier transform (FFT) to solve problems, utilizing matrix transposes for data permutation that often requires all-to-all communication. We use a discrete 3D FFT PDE solver in the NPB~\cite{Bailey1991} benchmark suite.
\item N-Body Methods simulate interactions among particles and are often compute-bound due to the high computational complexity. We use the BARNES benchmark from SPLASH~\cite{Woo1995,Sakalis2016}, an implementation of the Barnes--Hut method.
\item Structured Grids use stencil operations on regular grid structures. We use a PCG solver with a symmetric SMG preconditioner in the Hypre~\cite{Falgout2002} library for evaluation.
\item Unstructured Grids use irregular grid structures, and operations often involve multiple levels of memory references. We use OpenFOAM~\cite{Jasak2007}, a production CFD code implementing the finite volume method.
\item Monte Carlo methods rely on random trials to find approximate solutions. We use XSBench~\cite{Tramm2014}, a Monte Carlo neutron transport proxy application.
\end{itemize}

\section{Evaluation}
We begin this section by profiling the memory usage and access patterns of the seven HPC applications and six graph applications.
We then proceed by emulating three configurations of the CXL-enabled memory subsystem in Figure~\ref{fig:arch} to evaluate the performance impact on the workloads.

\subsection{Characterization of Memory Usage}\label{sec:memusage}
We evaluate three metrics of memory usage in this section -- \textit{capacity, page access, and bandwidth}. While capacity usage has been extensively studied in multiple large-scale studies on leadership clusters~\cite{peng2020memory,michelogiannakis2022case,peng2021holistic}, the other two metrics on page and bandwidth usage are important for understanding the need for memory resources and optimization opportunities.
\begin{figure}[bt]
\centering
\begin{minipage}{.9\linewidth}
\includegraphics[width=\linewidth]{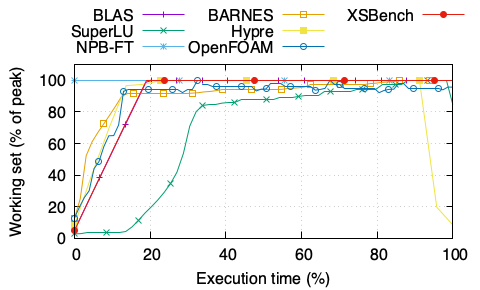}
\caption{Temporal profile of memory capacity usage in scientific workloads.}
\label{fig:memuse-hpc}
\end{minipage}\qquad
\begin{minipage}{.9\linewidth}
\includegraphics[width=\linewidth]{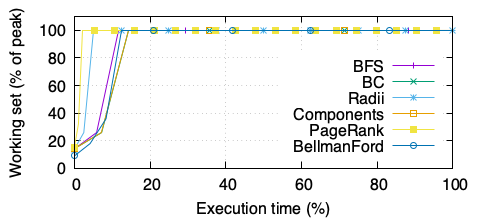}
\caption{Temporal profile of memory capacity usage in graph workloads.}
\label{fig:memuse-graph}
\end{minipage}
\end{figure}

\medskip\noindent\textbf{Dynamic Capacity Usage}
We quantify the capacity usage by summing up the resident set size (RSS) of each process during execution. To ease comparison across different applications, we normalize the capacity usage by the peak usage in each execution.
The results for scientific and graph workloads are shown in Figures~\ref{fig:memuse-hpc} and Figure~\ref{fig:memuse-graph}, respectively. We note that this RSS-based method is accurate for those single-process workloads, as well as Hypre which does not have a significant amount of shared pages. For OpenFOAM, however, this metric overestimates the total memory capacity by about $8\%$ because part of the working set is shared among all the MPI processes. A more accurate metric could use the proportional resident set size (PSS) that is available from \verb|smaps_rollup|, which accounts for shared pages.

After the initialization phase, we observe that the memory usage remains largely the same for the evaluated workloads. In BLAS, OpenFOAM, and Ligra, a warm-up phase is included in the initialization phase and then memory usage remains nearly constant during the computation phase. OpenFOAM has a small periodic variance in its iterative time steps. The memory usage of BARNES and Hypre grows during their initial computation phase due to the lack of warm-up in their initialization. The memory usage of SuperLU grows during the execution as the $L$ and $U$ matrices are filled in.

We note that more varied temporal capacity usage in jobs has been identified on large-scale production clusters~\cite{peng2021holistic}, which often involves complex workloads and coupled simulations. Based on the memory usage profile in this work, we determine that a static on-demand memory capacity allocation is suitable for all the evaluated workloads, e.g., statically compose the memory to match their peak usage, since their capacity usage has low variance over time.
 
\begin{figure}[bt]
\begin{minipage}[t]{\linewidth}
\vspace{0pt}
\includegraphics[width=0.49\linewidth]{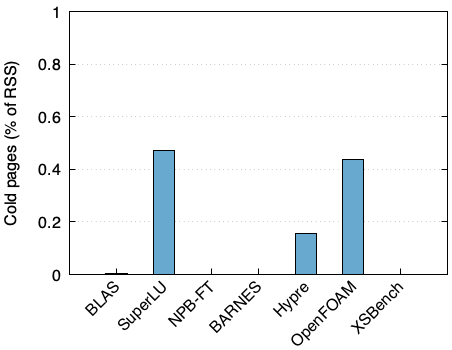}
\includegraphics[width=0.49\linewidth]{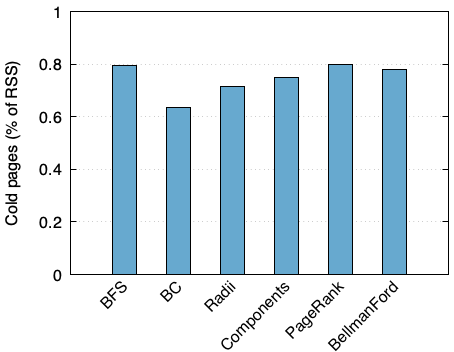}
\caption{The number of cold pages in scientific (left) and graph (right) workloads.}
\label{fig:memcold}
\end{minipage}\hfill
\begin{minipage}[t]{\linewidth}
\vspace{0pt}
\includegraphics[width=0.49\linewidth]{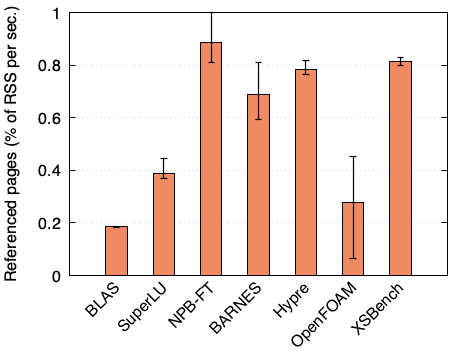}
\includegraphics[width=0.49\linewidth]{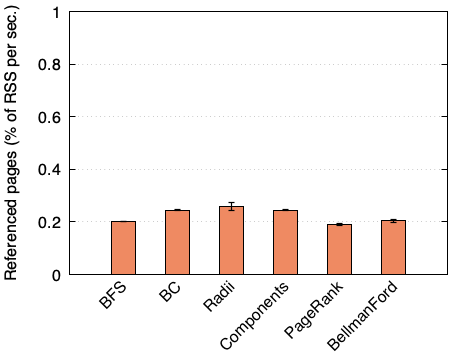}
\caption{The number of unique pages accessed over time in scientific (left) and graph (right) workloads. The bar reports the time average with quartiles.}
\label{fig:memvar}
\end{minipage}
\end{figure}

\medskip\noindent\textbf{Dynamic Page Access Patterns} A common optimization strategy in heterogeneous memory systems is to place less frequently accessed pages on slower memory tiers. To understand if this approach is applicable, we measure the number of cold pages in the evaluated workloads. Similar to \cite{Maruf2022}, we define a page as cold if it is never referenced in the entire computation phase, i.e., from the end of an application's initialization phase until the end of its computation phase. We note that this is a ``extreme" definition of cold pages as many existing works separate relative hot and cold pages for optimization, which is beyond the scope of this work. The data is measured by counting the per-page Accessed bits, similar to \cite{agarwal2017thermostat}, using our profiler in interrupt mode and normalized by the final RSS.

Figure~\ref{fig:memcold} shows the amount of cold pages in each workload. Four out of the seven scientific workloads have no significant number of cold pages at all. This is to be expected for many common HPC patterns that use all of their input data and have known output sizes. In SuperLU, the size of the output is not known a priori, which may cause some amount of cold memory due to oversized buffers. In contrast, the amount of cold memory is high for all of the graph workloads. One possible reason is that graph processing frameworks often allocate auxiliary data structures to optimize performance.

Although our definition of cold page is coarse-grained,
the results indicate that page placement techniques based on cold pages are not applicable to many HPC workloads, where most of the memory is used during execution. However, cold memory is present in the SuperLU, Hypre, and OpenFOAM workloads. More work is required to understand the cause of the cold pages to determine if such classification-based page placement is possible.

To identify memory pages that are only occasionally accessed, a more fine-grained metric such as the average re-use interval could be used \cite{lagar2019software}. Such per-page or per-cacheline tracking requires a more elaborate profiling method with higher privileges, it is not possible using \texttt{smaps} alone.

\medskip\noindent\textbf{Dynamic Bandwidth Usage} 
\begin{figure}[bt]
\centering
\includegraphics[width=\linewidth]{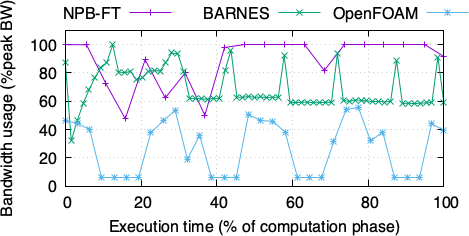}
\caption{Varied memory bandwidth usage over time in three scientific applications.}
\label{fig:memprof-time}
\end{figure}
We approximate the dynamic memory bandwidth usage in the workloads in two steps. First, we identify the number of unique pages accessed in every one-second interval and present the average. Then, for applications that exhibit high variance, we further present the measured bandwidth usage over time. We note that this estimation of bandwidth usage could be lower than the actual bandwidth as it averages over the one-second time period. This interval was found to be sufficient for our workloads and is finer than the intervals used in previous works~\cite{Maruf2022}. Also, the assumption is that the whole page will be transferred from the memory controller, while sparse access to pages may result in a few bytes being transferred. More precise fine-grained profiling will track traffic to and from the memory controller, which is part of our future work. 

Figure~\ref{fig:memvar} shows that BLAS and Ligra have only around 20\% of the working set referenced in each interval. In contrast, other scientific applications reference almost all pages in each interval. For instance, NPB-FT shows the highest referenced pages per interval due to the frequent use of matrix transposition in the fast Fourier transform. 

We choose three scientific applications that exhibit significant time-variance in their page access in Figure~\ref{fig:memvar} to further profile their dynamic memory bandwidth usage. Figure~\ref{fig:memprof-time} presents the varied memory bandwidth usage over their execution phase. The variability for NPB-FT and BARNES can primarily be attributed to warm-up during the first part of the compute phase, which was previously indicated by Figure~\ref{fig:memuse-hpc}. The results for OpenFOAM and BARNES show a clear variation in the memory usage pattern during different phases of each time step.

\subsection{Composable Memory Capacity\label{sec:cap}}
\begin{figure}[bt]
\begin{minipage}[t]{0.5\linewidth}
\vspace{0pt}
\includegraphics[width=\linewidth]{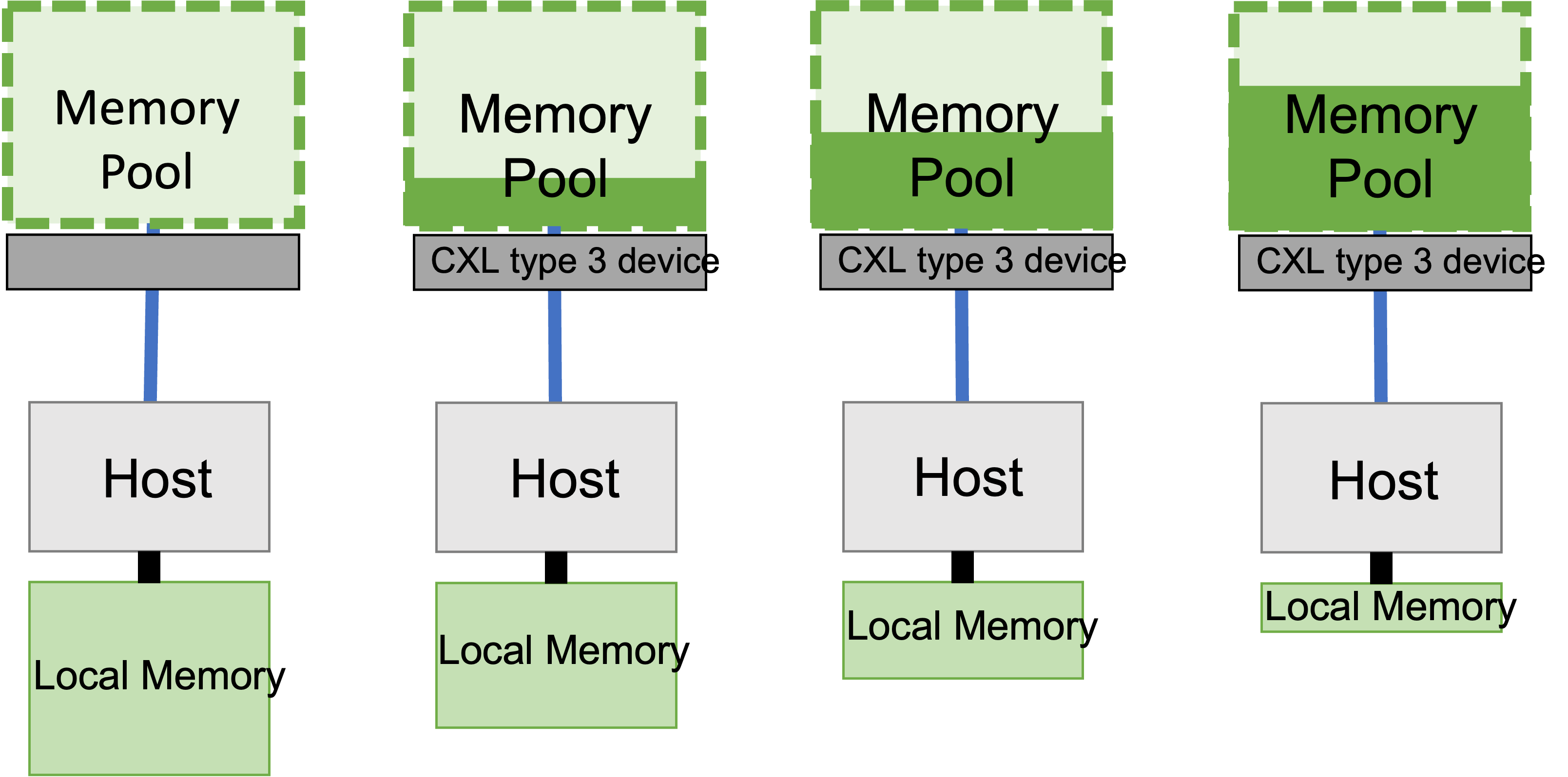}
\end{minipage}\hfill
\begin{minipage}[t]{0.45\linewidth}
\vspace{20pt}
\resizebox{\linewidth}{!}{
\begin{tabular}[t]{@{} l r r @{}}
    \toprule
    Configuration  & Bandwidth & Latency \\ \midrule
    Local Memory & 49 GB/s & 113 ns \\
    Memory Pool  & 28 GB/s & 205 ns \\
    \bottomrule
   \end{tabular}
   }
\end{minipage}
\caption{Four compositions of the memory subsystem using a variable amount of local memory and pooled memory.\label{fig:setup-capacity-scaling}}
\end{figure}
\begin{figure*}
\centering
\begin{minipage}[b]{.45\textwidth}
\includegraphics[width=\linewidth]{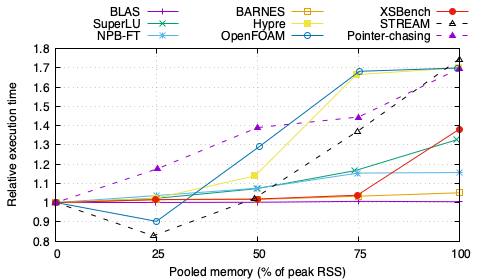}
\caption{The performance of seven scientific applications on five composed memory systems using an increased ratio of pooled memory. The performance is normalized to that on 100\% local memory.}
\label{fig:emu-hpc}
\end{minipage}\qquad
\begin{minipage}[b]{.45\textwidth}
\includegraphics[width=\linewidth]{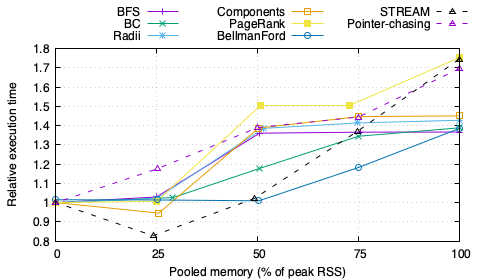}
\caption{The performance of six graph applications on five composed memory systems using an increased ratio of pooled memory. The performance is normalized to that on 100\% local memory.}
\label{fig:emu-ligra}
\end{minipage}
\end{figure*}

CXL-enabled memory pooling can enable fine-grained provisioning of memory capacity for applications. We evaluate the composability of a memory subsystem consisting of local memory and a memory pool at different sizes, illustrated in Figure~\ref{fig:setup-capacity-scaling}, by emulating it on the Intel testbed. The  memory pool has about 50\% bandwidth and 90 ns extra latency compared to the local memory. We measure the peak memory usage reported by the profiler to configure the composed memory subsystem so that 0\%, 25\%, 50\%, 75\%, and 100\% of each job's memory usage will be supported by memory pools. In this way, we emulate a potential use case on future HPC systems, where each job can compose its own memory subsystem dynamically.

We measure the execution time of the workloads on the emulated configurations to characterize the impact of memory pools on their performance. Figure~\ref{fig:emu-hpc} and Figure~\ref{fig:emu-ligra} report their relative performance compared to using only local memory. Additionally, we show the relative bandwidth and latency changes as measured by the STREAM and pointer-chasing benchmarks in each configuration. Note that the reported application performance is obtained without any optimizations in memory management or data partitioning, and extensive works on multi-tier memory systems have shown that further performance optimization can be achieved~\cite{dulloor2016data,wu2017unimem,agarwal2017thermostat,chen2020atmem}. We focus our analysis on three composition ratios -- 25\%, 50\%, 75\%, as the two systems that use 100\% local memory and 100\% pooled memory actually have lower aggregated memory bandwidth due to a reduced number of memory channels. 

In general, the HPC workloads exhibit better performance than the graph workloads on the memory systems composed of memory pools. With 75\% pooled memory, five out of seven HPC applications have less than 18\% performance degradation. In contrast, the graph workloads generally show higher sensitivity to an increased amount of pooled memory. At 75\% pooled memory, five out of six graph applications have a 35\%-50\% slowdown. This is expected since graph workloads have low operational intensity and are sensitive to increased latency.

The HPC workloads can be classified into three classes based on their performance at 75\% pooled memory.
\begin{itemize}[leftmargin=*]
 \item Class I: bandwidth insensitive. BLAS, BARNES, and XSBench show little performance changes at different configurations of memory pooling. This is expected for BLAS level 3 and BARNES which both have a high operational intensity. Also, the DGEMM operation in BLAS is highly optimized for cache locality and thus less sensitive to main memory. We expected XSBench to be sensitive to higher memory latency at 75-25\% local memory due to its random access pattern, however, this was not observed.
 \item Class II: bandwidth moderate. SuperLU and NPB-FT both show minor performance degradation (less than 15\%). Note that no optimizations such as paging migration and data placement are applied, which may further mitigate the performance impact. 
 \item  Class III: bandwidth sensitive. Only two out of seven HPC applications fall into this category -- OpenFOAM and Hypre. Both are highly sensitive to memory bandwidth due to low operational intensity and many indirect memory accesses. 
\end{itemize}

\subsection{Composable Memory Bandwidth}
\begin{figure}[bt]
\begin{minipage}[t]{0.5\linewidth}
\vspace{0pt}
\includegraphics[width=\linewidth]{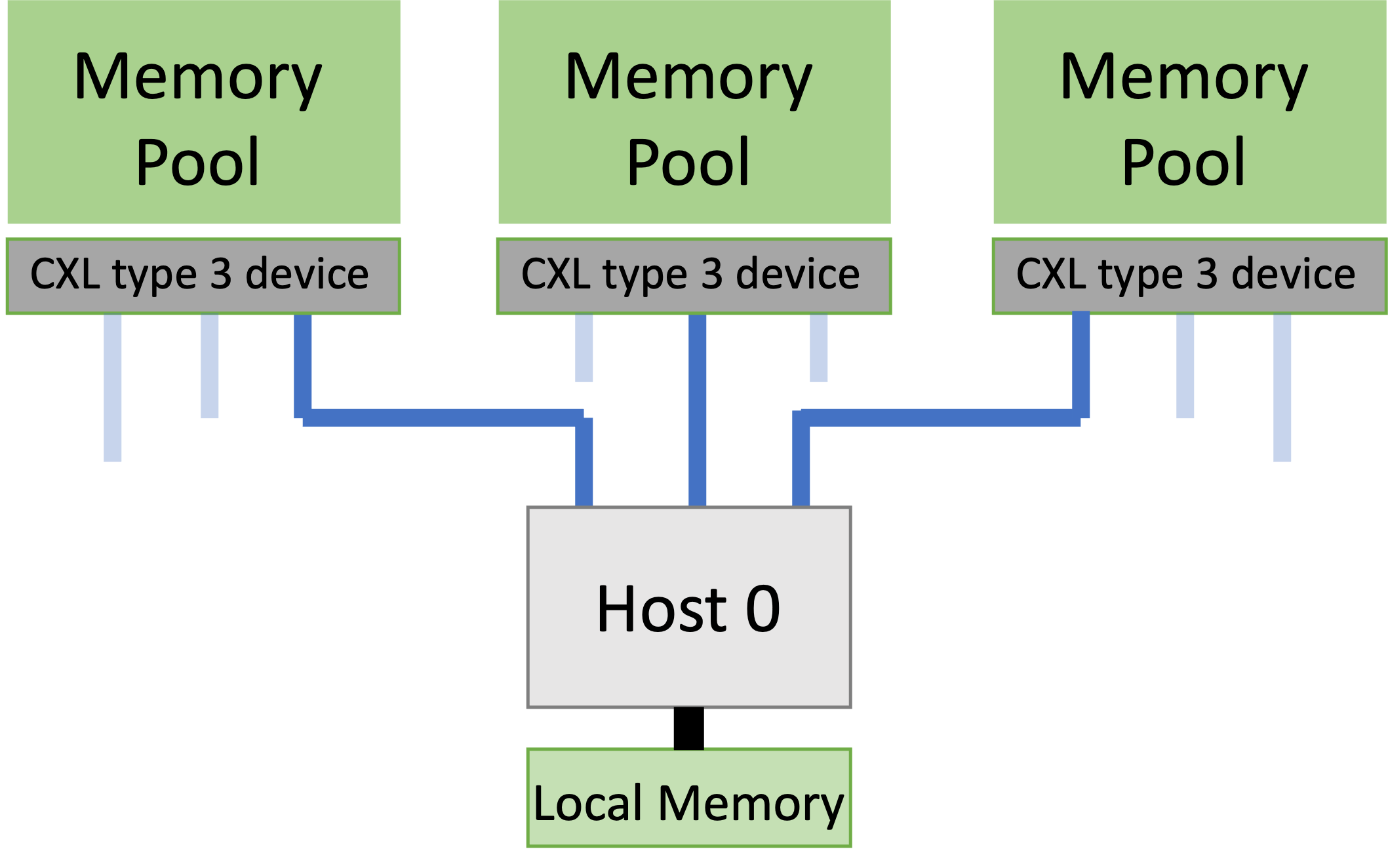}
\end{minipage}\hfill
\begin{minipage}[t]{0.4\linewidth}
\vspace{10pt}
\resizebox{\linewidth}{!}{
\begin{tabular}[t]{@{} l r r @{}}
    \toprule
    Configuration  & Bandwidth \\ \midrule
    Local Memory & 33 GB/s \\
    One Link & 32 GB/s \\
    Two Links  & 64 GB/s \\
    Three Links & 96 GB/s \\
    \bottomrule
   \end{tabular}
   }
\end{minipage}
\caption{An emulated high-bandwidth configuration of the memory system. We turn on an increased number of CXL links between the host and CXL type 3 devices. \label{fig:setup-bw-scaling}}
\end{figure}
\begin{figure}[bt]
\centering
\includegraphics[width=0.8\linewidth]{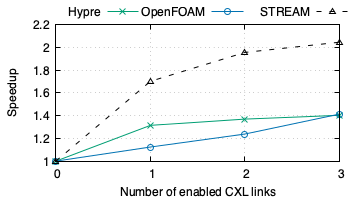}
\caption{Selected HPC workloads in bandwidth-scaled memory configuration.}
\label{fig:bwscale}
\end{figure}

The second use case for CXL-enabled memory pooling in HPC systems is to provide composable memory bandwidth. In this experiment, we show that a scalable high-bandwidth memory subsystem can be configured on the same system in Figure~\ref{fig:arch} to support bandwidth-intensive workloads. Currently, bandwidth-intensive workloads often need expensive HBM memory or large aggregated bandwidth on multiple nodes, leading to the under-utilization of memory capacity.

We emulate a scalable high-bandwidth system on the AMD testbed described in Section~\ref{sec:setup}. We use the workloads classified as bandwidth-sensitive from the evaluation in Section~\ref{sec:cap} (OpenFOAM and Hypre) as they are the only two that exhibit significant performance degradation on pooled memory. We run each application on NUMA node 0 and scale up the number of memory nodes in use. Each additional NUMA node represents an additional CXL link connected to a separate memory pool as illustrated in Figure~\ref{fig:setup-bw-scaling}. The testbed has four NUMA domains, each with a similar bandwidth such that the theoretical bandwidth scales from 33 GB/s (i.e., 0 CXL links, only local memory) to 129 GB/s (i.e., 3 CXL links and local memory). In this case, we use the interleaving NUMA allocation policy meaning that memory pages are allocated on each available node in a round-robin fashion.
Interleaving maximizes the potential bandwidth advantage of using multiple memory controllers.

Figure~\ref{fig:bwscale} reports the relative performance compared to execution on the system with only local memory (i.e., no CXL-connected memory pools).
The results show that both Hypre and OpenFOAM can benefit from the increased memory bandwidth, with up to 40\% speedup at four emulated CXL links. This indicates that bandwidth scaling may be a promising use case of CXL-enabled memory for some HPC workloads. For OpenFOAM, the scaling is linear in the number of nodes, while for Hypre most of the speedup is achieved already at two nodes. A possible explanation is that Hypre's bandwidth requirements are saturated already at two emulated CXL links.

The results show that CXL links could enable more scalable and cost-effective high-bandwidth system configurations. In contrast, current systems fully equipped with HBM memory are known to be expensive and cannot support decoupling between capacity and bandwidth, which is essential for enabling high utilization.

\subsection{Interference on Shared Memory Pooling}
In practical use cases on HPC systems, multiple hosts share a memory pool. Depending on the workloads and memory pool characteristics, sharing may lead to performance degradation due to contention. Also, bulk-synchronous MPI applications may face challenges as many processes operate in a synchronized fashion so burstiness in memory requests to memory servers may cause periodic strong interference on shared memory devices. 

\begin{figure}[bt]
\begin{minipage}[t]{0.45\linewidth}
\vspace{0pt}
\includegraphics[width=\linewidth]{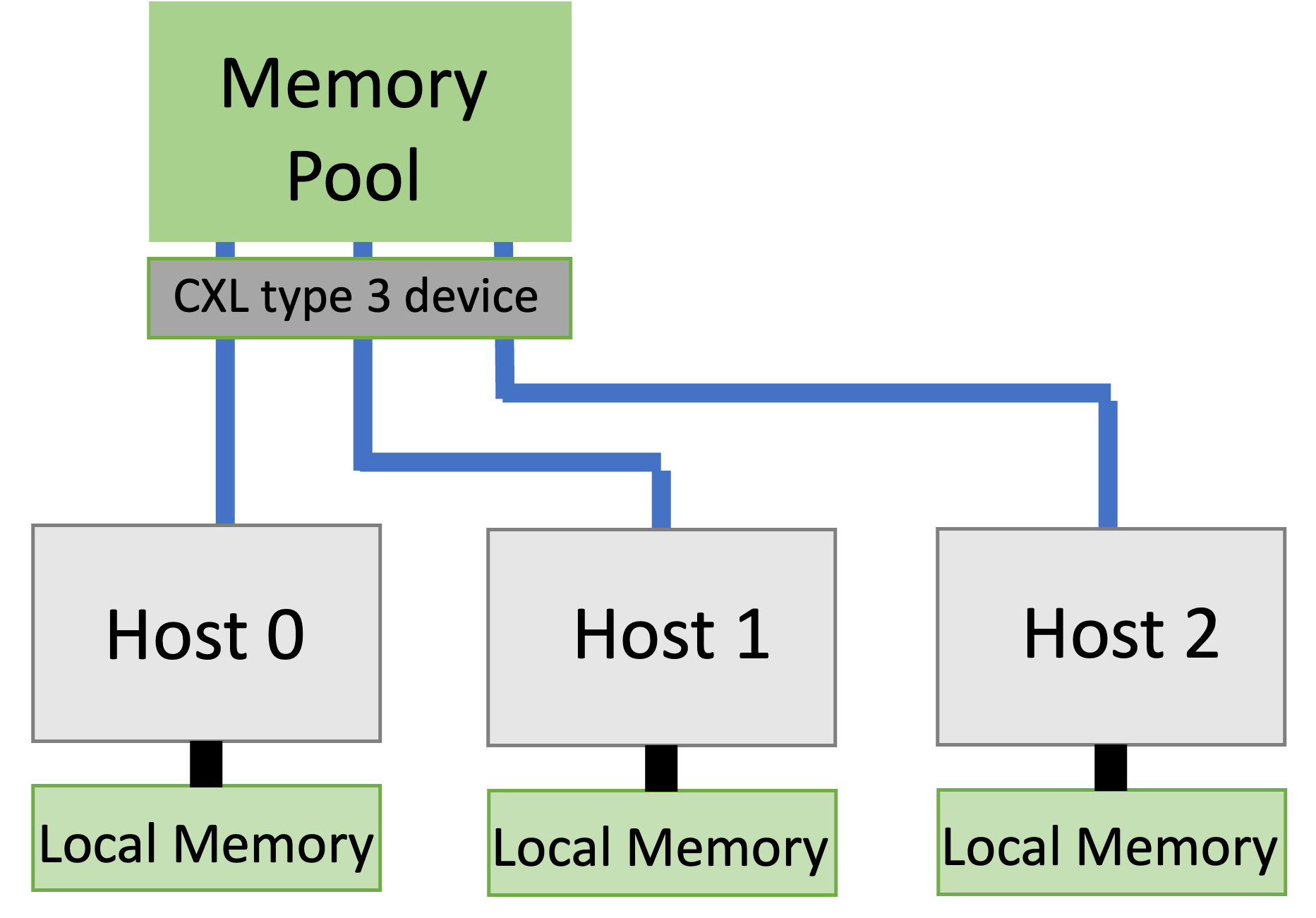}
\end{minipage}\hfill
\begin{minipage}[t]{0.4\linewidth}
\vspace{10pt}
\resizebox{\linewidth}{!}{
\begin{tabular}[t]{@{} l r r @{}}
    \toprule
    Configuration  & Bandwidth\\ \midrule
    Peak Memory Pool  & 33 GB/s \\
    Single Link  & 32 GB/s \\
    Two shared Links  & 16.5 GB/s \\
    Three shared Links  & 11 GB/s \\
    \bottomrule
   \end{tabular}
   }
\end{minipage}
\caption{An emulated configuration of a memory pool shared by multiple hosts. We increase the number of shared hosts from one to three, evaluating the impact of memory sharing on applications.\label{fig:sharing}}
\end{figure}

\begin{figure}[bt]
\begin{subfigure}[t]{\linewidth}
\includegraphics[width=\linewidth]{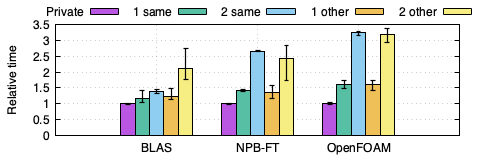}
\caption{HPC workloads.}
\label{fig:sharing-hpc}
\end{subfigure}\hfill
\begin{subfigure}[t]{\linewidth}
\includegraphics[width=\linewidth]{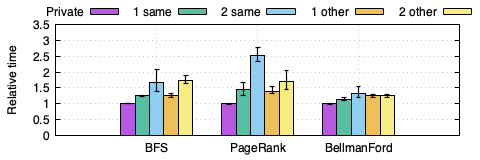}
\caption{Graph workloads.}
\label{fig:sharing-graph}
\end{subfigure}
\caption{The average execution time of different workloads when sharing a memory pool with zero, one, or two other hosts. Private means that the pool is not shared, "same" denotes that the other hosts run the same application, while "other" denotes that other hosts run different HPC applications. For instance, "BLAS 2 other" shows the execution time of BLAS when sharing with one host running NPB-FT and another host running OpenFOAM. "OpenFOAM 1 other" is the average execution time of OpenFOAM when sharing with another host running either BLAS or NPB-FT.}
\label{fig:sharing-both}
\end{figure}

We design a set of experiments to evaluate the effect of memory sharing in bulk-synchronous applications, as well as in unrelated jobs. 
On the AMD testbed, we use NUMA nodes 0-2 to represent three hosts, and NUMA node 3 to represent a shared memory pool. The setup is illustrated in Figure~\ref{fig:sharing}. The hosts use interleaving memory allocation to distribute the working set equally over the local and pooled memory.
On this system, we evaluate one HPC application from each of the three classes identified in Section~\ref{sec:cap}. We also select three different graph applications. We measure the execution time of the workloads on host 0 while varying the workloads on the other hosts (chosen among the same three applications).

The average execution times are presented in Figure~\ref{fig:sharing-hpc} and Figure~\ref{fig:sharing-graph}, with whiskers indicating the min and max time. In general, the execution time of all workloads increases as we increase the number of hosts using the same memory pool. As expected from the previous sensitivity analysis, among the HPC workloads BLAS is the least affected while OpenFOAM is the most affected. For NPB-FT and OpenFOAM, the performance degradation is substantial -- with just two hosts the slowdown is around 50\%, and with three hosts the slowdown is over 2x. This is in line with the reduction in bandwidth per host as we increase the number of enabled CXL links. For BLAS, the slowdown is moderate even with three hosts. However, if BLAS shares the memory pool with two hosts running more memory-intensive workloads, its performance is degraded substantially too.

The results show that the bandwidth provided by the emulated system when shared by multiple hosts is insufficient to support class II and III applications simultaneously. It is possible to support them if higher bandwidth links can be provided. Conversely, class I applications may run on a relatively simple shared memory pool. However, even class I applications may suffer severe degradation if sharing with more intensive workloads.

In general, the impact of sharing is less severe for the graph workloads. Compared to the other workloads, BellmanFord has a very low sensitivity to pool sharing in all configurations, and may thus run on simple shared memory pools. This correlates well with the results in Section~\ref{sec:cap}, which showed that the performance of BellmanFord is unaffected at 50\% pooled memory. On the shared system, BFS also has acceptable performance as long as only two hosts are used. The performance of PageRank degrades more when sharing the pool with other PageRank workloads rather than different types of graph workloads, as the other two graph workloads are relatively undemanding for the memory pool. 

These results show that performance may be severely degraded if memory pools become overloaded. Therefore, the job scheduler in an HPC system with CXL-enabled memory pools must take into account the dynamic memory usage profiles of jobs in Section~\ref{sec:memusage} to prevent overloading shared resources among jobs with conflicting demands.

\section{Related Works}
Network-attached memory was the main option for memory disaggregation before the emergence of CXL~\cite{Lim2012, gu2017efficient, peng2020memory}. These designs rely on swapping memory pages between the local and far memory across the network and often use RDMA for performance. A recent work~\cite{Gouk2022} compared network-attached memory with the load/store interface offered by CXL. They used a hardware prototype based on FPGAs and customized CPUs to show that CXL can improve performance on database and graph workloads.

An FPGA prototype is also used in~\cite{Maruf2022} to measure the performance impact of CXL-enabled memory on common data center workloads. The authors propose a page migration mechanism to minimize the performance degradation, based on the observation that the data center workloads have a significant portion of cold memory. Unlike our method, the page temperature characterization is based on PEBS monitoring in Intel CPUs, which requires the \verb|perf_events| subsystem to be enabled. This means it is not usable in many production HPC systems. Recently, Samsung has published performance results for the first ASIC-based CXL memory device \cite{park2022scaling}. Their results show better performance compared to FPGA prototypes and within 10\% of local DDR memory. They also demonstrate how the device can be used to scale out memory bandwidth for machine learning applications.

Leveraging NUMA to emulate CXL-enabled memory has previously been used to study the latency sensitivity of mainly data center workloads~\cite{Li2022}. However, the presented emulation method relies on kernel-level NUMA configuration, requiring root privileges. The authors propose a technique to predict the latency sensitivity and amount of cold memory in a workload to balance the amount of local and remote memory while minimizing performance degradation.

Alternatively, another memory system emulation technique is the use of the Linux \verb|resctrl| interface to configure memory bandwidth throttling in x86 CPUs~\cite{Foyer2021}. The authors study memory bandwidth sensitivity in a set of HPC workloads. However, the method is less suited to study CXL-enabled memory since it primarily affects bandwidth and not latency. It also lacks a simple way to emulate a system with several different types of memory, e.g. both local and remote memory.

Future disaggregated memory systems offering peer-to-peer communication (e.g. CXL 3.0 fabric) also enable new parallel computing models, for instance in deep learning training~\cite{Wang_2022}. In this paper, the authors propose a decentralized parameter communication scheme in a system with interconnected memory devices. The scheme provides improved performance over a conventional centralized scheme by reducing communication overheads. In distributed graph analytics, FAM-Graph~\cite{zahka2022fam} can significantly reduce memory usage by storing read-only data in a shared memory pool. In this work, remote memory is emulated using RDMA with an object-based get/put model.

Disaggregated memory is a kind of multi-tier memory or heterogeneous memory, where
data placement is an active research direction. Specific optimizations for KV-store, in-memory database, and graph analytics workloads have been proposed for DRAM-NVM systems~\cite{dulloor2016data, agarwal2017thermostat, chen2020atmem, Maruf2022}. For instance, Thermostat~\cite{agarwal2017thermostat} uses TLB-poisoning and a custom kernel module to estimate the access rate of a process's working set at runtime and then separate pages in different memory tiers.

\section{Discussion and Conclusions}
New workloads on HPC systems exhibit increasingly diverse requirements on the memory subsystem. Also, machine learning and data analytics have different limiting factors regarding memory bandwidth and capacity. Furthermore, complex workflows composed of these characteristically different components are also emerging. 

In this work, we focused on an HPC memory subsystem design comprising type 3 CXL devices. We show that such a system could enable dynamic provisioning of memory capacity and bandwidth scaling that better matches each application's usage. In particular, we provide an emulator and profiler that can quickly estimate a target application's potential impact on a CXL-enabled memory system. Our results show that five out of seven evaluated scientific applications have little performance degradation on an emulated CXL system when memory pools back 75\% of their memory usage. Furthermore, on an emulated high-bandwidth CXL system, even the most bandwidth-sensitive applications like OpenFOAM and Hypre sustained their performance, highlighting the potential of CXL-enabled memory systems for scalable, cost-effective high-bandwidth systems. We also identified that the interference on shared memory servers needs to be addressed by coordination at the system level, based on an understanding of each job's dynamic usage, to mitigate performance degradation.

\section*{Acknowledgment}
The experiments were enabled by resources provided by the Swedish National Infrastructure for Computing (SNIC). This work was partially performed under the auspices of the U.S. Department of Energy by Lawrence Livermore National Laboratory under contract No. DE-AC52-07NA27344. LLNL-CONF-838711.

\bibliography{main}

\end{document}